\documentclass[12pt,onecolumn]{IEEEtran}

\ifCLASSINFOpdf
\else
\fi

\usepackage{graphicx,amssymb,amsmath}
\usepackage{multicol}
\usepackage[noadjust]{cite}
\usepackage{setspace}
\usepackage{stfloats}
\usepackage{cite}
\usepackage{amsthm}
\usepackage{flushend}
\usepackage{cases,subeqnarray}
\usepackage{bm,multirow,bigstrut}
\usepackage{textcomp}
\usepackage{latexsym,bm}
\usepackage{xcolor}
\usepackage{mathtools}
\usepackage{dsfont}
\usepackage{extarrows}
\usepackage{algorithm}
\usepackage{algpseudocode}
\usepackage{amsmath}
\usepackage{booktabs}
\newcommand{\tabincell}[2]{\begin{tabular}{@{}#1@{}}#2\end{tabular}}
\usepackage{multicol}       
\usepackage{multirow}       
\usepackage{array}          
\usepackage{makecell}

\theoremstyle{plain}

\theoremstyle{plain}

\begin{document}

\title{Improving Sum-Rate of Cell-Free Massive MIMO with Expanded Compute-and-Forward}

\author{Jiayi Zhang,~\IEEEmembership{Senior Member,~IEEE,}
        Jing Zhang,
        Derrick~Wing~Kwan Ng,~\IEEEmembership{Fellow,~IEEE,}\\
        Shi Jin,~\IEEEmembership{Senior Member,~IEEE,} and Bo Ai,~\IEEEmembership{Senior Member,~IEEE}
\thanks{J. Zhang and J. Zhang are with the School of Electronic and Information Engineering, Beijing Jiaotong University, Beijing 100044, China, and also with the Frontiers Science Center for Smart High-speed Railway System, Beijing Jiaotong University, Beijing 100044, China (e-mail: jiayizhang@bjtu.edu.cn).}
\thanks{D. W. K. Ng is with the School of Electrical Engineering and Telecommunications, University of New South Wales, NSW 2052, Australia. (e-mail: w.k.ng@unsw.edu.au).}
\thanks{S. Jin is with the National Mobile Communications Research Laboratory, Southeast University, Nanjing 210096, China (e-mail: jinshi@seu.edu.cn).}
\thanks{B. Ai is with the State Key Laboratory of Rail Traffic Control and Safety, Beijing Jiaotong University, Beijing 100044, China, and also with the Frontiers Science Center for Smart High-speed Railway System, and also with Henan Joint International Research Laboratory of Intelligent Networking and Data Analysis, Zhengzhou University, Zhengzhou 450001, China, and also with Research Center of Networks and Communications, Peng Cheng Laboratory, Shenzhen, China (e-mail: boai@bjtu.edu.cn).}}


\maketitle

\begin{abstract}
Cell-free massive multiple-input multiple-output (MIMO) employs a large number of distributed access points (APs) to serve a small number of user equipments (UEs) via the same time/frequency resource. Due to the strong macro diversity gain, cell-free massive MIMO can considerably improve the achievable sum-rate compared to conventional cellular massive MIMO. However, the performance of cell-free massive MIMO is upper limited by inter-user interference (IUI) when employing simple maximum ratio combining (MRC) at receivers.
To harness IUI, the expanded compute-and-forward (ECF) framework is adopted.
In particular, we propose power control algorithms for the parallel computation and successive computation in the ECF framework, respectively, to exploit the performance gain and then improve the system performance.
Furthermore, we propose an AP selection scheme and the application of different decoding orders for the successive computation. Finally, numerical results demonstrate that ECF frameworks outperform the conventional CF and MRC frameworks in terms of achievable sum-rate.
\end{abstract}


%
\IEEEpeerreviewmaketitle

\section{Introduction}
\IEEEPARstart{M}{assive} multiple-input multiple-output (MIMO) is a promising physical-layer technology to keep up with the exponential traffic growth of future wireless communication systems.
More specifically, massive MIMO can provide tremendous beamforming gains and spatially multiplexing gains to multiple user equipments (UEs) and increase the system achievable sum-rate \cite{marzetta2010noncooperative,zhang2020prospective,chen2020massive}.
Despite the potential performance gain brought by massive MIMO, UEs at cell-edge may experience poor channel conditions and suffer from strong inter-cell interference (ICI).
To alleviate this performance bottleneck, distributed massive MIMO has been proposed to combat ICI and to improve the performance of cell-edge UEs.
However, there is a fundamental performance limitation for distributed massive MIMO with full cooperation between different transmitters \cite{lozano2013fundamental}.

Recently, the authors in \cite{ngo2017cell} proposed a practical network infrastructure for distributed massive MIMO, under the name of cell-free massive MIMO \cite{nayebi2017precoding,karlsson2019techniques,bashar2019on}.
In cell-free massive MIMO systems, a large number of access points (APs) distribute in a large area and are connected to a central processing unit (CPU) via a fronthaul network.
In particular, a small number of UEs are served by all APs with the same time/frequency resource \cite{ngo2017cell, interdonato2019ubiquitous, bjornson2020making}.
Since there are no cells or cell boundaries, ICI does not exist.
Indeed, cell-free massive MIMO is a specific realization of distributed massive MIMO \cite{ngo2017cell}.

The most outstanding aspect of cell-free massive MIMO is that many APs simultaneously serve a much smaller number of UEs, which yields a high degree of macro-diversity and can offer a huge spectral efficiency.
Besides, some studies have reported that favorable propagation is also a potential advantage for cell-free massive MIMO which can be exploited to eliminate inter-user interference (IUI) \cite{ngo2017cell}.
Note that favorable propagation refers to the property that when the number of AP antennas is sufficiently large, the channels between the UEs and APs become asymptotically orthogonal \cite{bjornson2017massive}.
However, the favorable propagation property does not always hold in practical systems.
The non-negligible IUI is highly undesirable and leads to a considerable loss in achievable sum-rate.
As a result, how to harness the IUI has triggered many new coding and signal processing techniques.

\subsection{Related Works}
As a new approach of linear physical-layer network coding that allows intermediate nodes to send out functions of their received packets \cite{ahlswede2000network,li2003linear,yang2014optimal,nazer2011reliable}, the compute-and-forward (CF) scheme has recently been employed in cell-free massive MIMO systems to offer protection against noise and to reduce IUI with cooperation gain \cite{huang2017compute}.
For the uplink transmission, UEs employ a nested lattice coding strategy to encode data that takes values in a prime-size finite field before transmission.
Then, the CF scheme enables APs to decode the integer linear equations of UEs' codewords using the noisy linear combinations provided by the channels.
Relying on nested lattice codes, the linear combination of UEs' codewords is still a regular codeword \cite{nazer2011compute,hong2013compute}.
Next, each AP forwards the decoded combination to the CPU through the fronthaul link.
After receiving sufficient linear combinations, the CPU could recover every UE¡¯s original data by performing AP selection and solving the received equations \cite{feng2013algebraic,nokleby2011lattice, huang2017compute}.

However, the CF scheme requests all UEs transmit with equal power, which is generally not the optimal strategy for improving the achievable sum-rate.
Due to the different propagation conditions between APs and UEs, the performance can be improved by performing appropriate power control \cite{bjornson2017massive}.
Moreover, with power control for UEs, the effective noise variance across all APs whose linear combinations involve the message can be reduced.
Then, the achievable sum-rate can be further improved.

Motivated by the discussion above, we adopt the expanded compute-and-forward (ECF) framework which was proposed in \cite{nazer2016expanding} for the uplink transmission in cell-free massive MIMO systems.
The ECF framework is able to distribute transmit powers unequally and retains the connection between the finite field data and the lattice codeword.
We note that coordinated multiple points (CoMP) framework also can be implemented with interference alignment at the transmitter-side \cite{li2019coordinated,marsch2011coordinated,veeravalli2018interference}, however, the distinction between CoMP and ECF is that CoMP as conventionally defined does not involve CF strategy.

There are two types of ECF framework, named parallel computation and successive computation, respectively.
The distinction between these schemes is that in parallel computation the CPU recovers UEs' data independently while for successive computation the CPU decodes the linear combinations by using successive cancellation.
Specifically, in successive computation, the combinations which have been decoded can be used as side information in the subsequent decoding steps to decrease both effective noise variance and the number of UEs that need to tolerate the effective noise.
Applying successive computation helps improve the achievable sum-rate, however, in terms of processing delay, the parallel computation has some advantages. In other words, there is a trade-off between the parallel computation and successive computation.

Besides, there are some key aspects which dominate the performance of ECF framework: coefficient vector selection and AP selection.
Since the performance of ECF is captured by the computation rate and that rate achieves the highest when the equation coefficients closely approximate the effective channel coefficients, designing the coefficient vector elaborately is beneficial for the improvement of achievable sum-rate.
As for AP selection, it is performed at the CPU when recovering UEs' original data in both parallel computation and successive computation.
With the help of AP selection, the computational complexity of power optimization is reduced.
Furthermore, the noise tolerance on UEs' data can also be relaxed, which contributes to the improvement of the achievable sum-rate.

\subsection{Contributions}
In this paper, we consider the application of ECF framework in cell-free massive MIMO systems to increase the achievable sum-rate, including both parallel computation and successive computation.
The main contributions of this paper are as follows:
\begin{itemize}
  \item We apply a quadratic programming relaxation based coefficient vector selection method and a large-scale fading based low-complexity AP selection algorithm to improve the achievable sum-rate of the cell-free massive MIMO system.
  \item We design efficient power control algorithms for parallel and successive computation schemes, respectively.
        For the successive computation scheme, we further derive a sub-optimal decoding order of combinations and develop three assignment algorithms to find a sub-optimal decoding order of UEs.
  \item We quantitatively compare the performance of conventional combining and ECF frameworks under practical channel model and scenarios, which proves that the ECF framework is an effective approach for the fronthaul reduction.
      In particular, the successive computation scheme outperforms the parallel computation scheme with a larger fronthaul load.
\end{itemize}

Compared with our related conference paper \cite{zhang2019expanded}, which focused only on parallel computation with power control based on uplink-downlink duality, in this paper, we provide a thorough analysis for the successive computation scheme with power control for improving the achievable sum rate.
Besides, the problem-solving methodology for determining the suboptimal decoding order of combinations and UEs are investigated.
Furthermore, the results from \cite{zhang2019expanded} are not applicable to the case considered in this paper due to different power control method and additional AP selection algorithm are applied.
More importantly, we also provide practice insights into the performance of MRC, CF, centralized MMSE, parallel computation, and successive computation schemes in achievable sum rate.

The rest of this paper is organized as follows.
In Section \ref{system_model}, we describe the cell-free massive MIMO system model.
A detailed introduction for ECF framework is given in Section \ref{Expanded Compute-and-Forward}.
Furthermore, AP selection methods and power control algorithm for parallel computation are introduced in Section \ref{Parallel Computation}.
In Section \ref{Successive Computation}, we investigate different decoding order methods of combinations for successive computation.
Finally, numerical results and discussions are given in Section \ref{numerical_results} while Section \ref{conclusion} concludes the paper.

Table \ref{t1} shows the notations.
Unless further specified, plain letters, boldface letters, and boldface uppercase letters denotes scalars, column vectors, and matrices respectively.

\begin{table}[t]
\centering
\caption{Notations}
\begin{tabular}{|c|c|}
\hline
$p$ & A prime number\\\hline
$\mathbb{R}$, $\mathbb{C}$,  $\mathbb{Z}_p$ & \tabincell{c}{Reals, complex field, \\ finite field of size $p$}\\\hline
${q_1},{q_2},{w_1},{w_2}, r$ & Element in $\mathbb{Z}_p$ \\\hline
\tabincell{c}{$\mathbb{Z}[i] =$ \\ $\left\{ {a + \left. {bi} \right|a,b \in \mathbb{Z}} \right\}$} &  \tabincell{c}{Set of Gaussian integers whose real  \\  and imaginary parts are both integers\\}\\\hline
$\sum$ & \tabincell{c}{Addition over the real \\ or complex field}\\\hline
$ \oplus $ & Addition over the finite field \\\hline
\textcolor{black}{$ a \bmod p = r$} & \tabincell{c}{Computing the remainder \\ of dividing $a$ by $p$} \\\hline
${q_1}{w_1} \oplus {q_2}{w_2}$ & $ {{q_1}{w_1} + {q_2}{w_2}} \bmod p $ \\\hline
$\left\| {\bf{a}} \right\|$ & 2-norm of vector $\bf{a}$ \\\hline
${{\bf{a}}^T}$, ${{\bf{a}}^H}$ & \tabincell{c}{Transpose of $\bf{a}$,\\ conjugate-transpose of $\bf{a}$} \\\hline
$\left\lfloor a \right\rfloor $ & Floor function of $a$ \\\hline
$\textbf{I}$ & Identity matrix \\\hline
\textcolor{black}{${\mathbb{E}}\left\{  {a}  \right\}$} & \textcolor{black}{Expectation of ${a}$} \\\hline
${\log ^ + }\left( a \right)$ & \tabincell{c}{$\max \left( {\log \left( a \right),0} \right)$, the log function\\  is with respect to base 2}\\\hline
\end{tabular}
\label{t1}
\end{table}

\section{System Model}\label{system_model}
We consider an uplink cell-free massive MIMO system.
$M$ single-antenna APs and $L$ ($M > L$) single-antenna UEs are randomly distributed in a wide geographical area \cite{ngo2017cell, interdonato2019ubiquitous, bjornson2020making}.
APs provide services for UEs via the same time/frequency resource.
In particular, each AP exchanges information with the CPU via fronthaul link.
As the practical number of APs is finite, we assume that the IUI can still have significant impact on the achievable sum-rate.

\begin{figure}[t]
\centering
\includegraphics[width=3.5in]{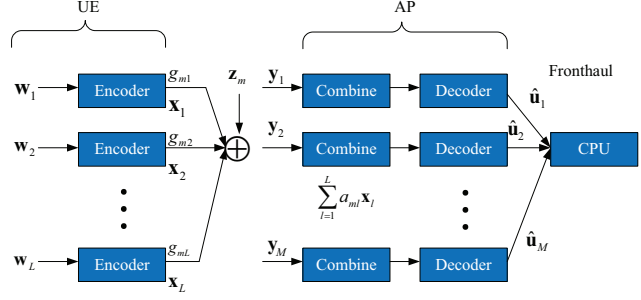}
\caption{ECF framework based cell-free massive MIMO systems.}
\label{model}
\end{figure}

First, we will provide some necessary definition on nested lattice codes.
An $n$-dimensional lattice, $\Lambda$, is a set of points in ${\mathbb{R}^n}$ such that if ${\mathbf{s}},{\mathbf{t}} \in \Lambda $, then ${\mathbf{s}} + {\mathbf{t}} \in \Lambda $ and if ${\mathbf{s}} \in \Lambda $, then $ - {\mathbf{s}} \in \Lambda $.
Note that a lattice can always be written in terms of a lattice generator matrix ${\mathbf{B}} \in {\mathbb{R}^{n \times n}}$, i.e., $\Lambda  = \left\{ {{\mathbf{s}} = {\mathbf{Bc}}:{\mathbf{c}} \in {\mathbb{Z}^n}} \right\}$.
Besides, a lattice $\Lambda$ is said to be nested in a lattice ${\Lambda _1}$ if $\Lambda  \subseteq {\Lambda _1}$.
As shown in Fig. \ref{model}, without loss of generality, the $l$th UE maps the original length-$k$ data ${\bf{w}}_l \in \mathbb{Z}^{k}_p$ into a length-$n$ complex-valued lattice codeword ${{\bf{x}}_l}$ with encoder ${\phi _l}:\mathbb{Z}_p^{k} \to \mathbb{Z}{[i]^n}$.
The specific choices of $n$ and $p$ are studied in \cite[Theorem 8]{nazer2016expanding}.
For creating generation matrices to encode the original data into nested lattice codeword, the blocklength needs to be large enough. Therefore, the longer blocklength, i.e., $n$, is better.
Note that $k_l$ is the number of symbols carrying information.
The remaining $k-k_l$ symbols are set to zero to meet the power constraint and the effective noise tolerance.
The lattice codeword is subject to the power constraint $\mathbb{E}{\left\| {{{\bf{x}}_l}} \right\|^2} \le n{P_l}$, where ${P_l}$ is the transmit power of the $l$th UE.

Let $g_{mk}$ represent the channel coefficient between the $m$th AP and $l$th UE, which is given by
\begin{equation}\label{g_ml}
{g_{ml}} = \beta _{ml}^{1/2}{h_{ml}},
\end{equation}
where $\beta _{ml}$ denotes the large-scale fading and $h_{ml} \in {\mathbb{C}}$ denotes the small-scale fading.
With the help of \cite[Eq. (17)]{bjornson2020making}, the propagation is given as
\begin{equation}
{\beta _{ml}}\left[ {{\rm{dB}}} \right] =  - 30.5 - 36.7{\log _{10}}\left( {{d_{ml}}/1{\text{m}}} \right) + {F_{ml}},
\end{equation}
where $d_{ml}$ represents the distances between the $m$th AP and the $l$th UE and ${F_{ml}} \sim {\mathcal{N}}(0,{4^2})$ is the shadow fading.
We assume that ${h_{ml}},m = 1, \ldots ,M,l = 1, \ldots ,L$, are independent and identically distributed (i.i.d.) ${\cal C}{\cal N}\left( {0,1} \right)$ \textcolor{black}{random variables} (RV)s.

The length-$n$ vector received signal at the $m$th AP is
\begin{equation}
{{\bf{y}}_m} = \sum\nolimits_{l = 1}^L {{g_{ml}}} {{\bf{x}}_l} + {{\bf{z}}_m},
\end{equation}
where the thermal noise ${{\bf{z}}_m} \in {\mathbb{C}^n}$ is elementwise independent and identically distributed (i.i.d.) $\mathcal{CN}\left( {0,{\sigma ^2}} \right)$.

The ECF framework manipulates the algebraic structure such that any Gaussian integer combination of lattice codewords is still a lattice point.
In cell-free massive MIMO, each AP endeavours to represent the received length-$n$ signal vector ${{\bf{y}}_m}$ with a Gaussian integer linear combination of UEs' codewords.
By applying an equalization factor $b_m$ and selecting the coefficient vector ${{\bf{a}}_m} = {\left[ {{a_{m1}},{a_{m2}}, \ldots ,{a_{mL}}} \right]^T} \in \mathbb{Z}{\left[ i \right]^L}$, the scaled received signal can be expressed as
\begin{equation}\label{eff_noi}
{b_m}{{\bf{y}}_m} = \sum\nolimits_{l = 1}^L {{a_{ml}}} {{\bf{x}}_l} + \underbrace {\sum\nolimits_{l = 1}^L {\left( {{b_m}{g_{ml}} - {a_{ml}}} \right)} {{\bf{x}}_l} + {b_m}{{\bf{z}}_m}}_{{\rm{effective\,noise}}}.
\end{equation}
Each AP is equipped with a decoder, ${\varphi _m}:\mathbb{Z}{\left[ i \right]^{{n}}} \to \mathbb{Z}_p^{{k}}$.
Then, AP decodes the received signal ${{\bf{y}}_m}$ into the finite field as ${{{\bf{\hat u}}}_m} = {\varphi _m} \left( {{{\bf{y}}_m}} \right)$, where ${{{\bf{\hat u}}}_m}$ is an estimation of the linear combination of original data ${{\bf{u}}_m} = \mathop  \oplus \limits_{l = 1}^L {q_{ml}}{{\bf{w}}_l} =  {\sum\nolimits_{l = 1}^L {{a_{ml}}{{\bf{x}}_l}} } \bmod p$.
\footnote{If the codeword spacing for a given data from the $l$th UE can tolerate the maximum effective noise across the APs whose linear combinations involve that data, the probability of decoding error is given as $\Pr \left( {\mathop  \cup \limits_{m = 1}^{{M_l}} \left\{ {{{{\bf{\hat u}}}_m} \ne {{\bf{u}}_m}} \right\}} \right) < \varepsilon $, where $M_l$ represents the number of APs whose combinations contain the data and $\varepsilon$ is a small positive number that tends to zero.}
The specific procedure for recovering messages for UEs is stated in \cite{nazer2011compute}.
Given $L$ linear combinations of messages with real and imaginary coefficient matrices ${{\bf{Q}}^R} = \left\{ {q_{ml}^R} \right\}$, ${{\bf{Q}}^I} = \left\{ {q_{ml}^I} \right\}$, the CPU can recover message ${{\bf{w}}_l}$ if there exists a vector ${\bf{c}} \in {\mathbb{Z}}_p^{M \times L}$ such that
\begin{equation}
{{\bf{c}}^T}\left[ {\begin{array}{*{20}{c}}
{{{\bf{Q}}^R}}&{ - {{\bf{Q}}^I}}\\
{{{\bf{Q}}^I}}&{{{\bf{Q}}^R}}
\end{array}} \right] = {{\bm{\delta}} _l^T},
\end{equation}
where ${{\bm{\delta}} _l}$ denotes a unit column vector with 1 in the $l$th entry and 0 elsewhere.
\footnote{The two decoders adopted at APs and the CPU, respectively, have different functionalities.
Indeed, decoders at the APs are used for decoding the received signal into the linear combination of the UEs' original data.
Then, these decoded linear combinations are transmitted from APs to CPU.
In contrast, the decoder at the CPU is responsible for recovering each UE's data from those combinations.
Specifically, when applying it with successive computation, the interference cancellation procedure takes place at the decoder at the CPU.}
For the traditional multiuser MIMO systems, where $M=L$, data recovery is a major challenge due to the high probability of rank deficiency.
However, the number of APs is far larger than that of the UEs in cell-free massive MIMO systems.
Since when the number of APs increases the probability of selecting $L$ APs that provides $L$ independent linear combinations also increases \cite{huang2017compute}, the extra APs can ensure a much higher probability for avoiding rank deficiency, so as to improve the probability to recover the desired message.

\section{Expanded Compute-and-Forward}\label{Expanded Compute-and-Forward}
One of the major challenges in cell-free massive MIMO is the IUI in the uplink.
In particular, CF scheme can achieve large gain through decoding linear functions of transmitted signals with nested lattice codes.
The performance of CF scheme for cell-free massive MIMO has been compared with MRC in \cite{huang2017compute}, which shows that with equal power transmission at all UEs, the CF scheme can offer a throughput improvement.
Furthermore, the ECF framework can improve the achievable sum-rate utilizing the characteristic of optimal power control.

In this section, two practical ECF frameworks are considered for cell-free massive MIMO systems.
The first one is parallel computation, which refers to that the CPU decodes each of the integer linear combinations independently. Furthermore, successive computation decodes received combinations one-by-one and employing the side information to reduce the effective noise. We begin with the parallel computation.

\subsection{Coefficient Vector Selection}
The goal of this paper is to evaluate the performance of the ECF framework for cell-free massive MIMO systems by deriving its computation rate region \cite{nazer2016expanding}, which is defined as the set of achievable rate $R_l$ ensuring successful data recovery:
\begin{align}
&{\mathcal{R}_{\text{ECF}}} \left({\bf{P}},{{\bf{g}}_m},{{\bf{a}}_m}\right) \triangleq  \bigg \{ \left({R_1},{R_2},\dots,{R_L}\right) \in \mathbb{R}_ + ^L:\notag\\
&{R_l} \! \le \!{\log ^ + }\!\left(\frac{{{P_l}}}{{{\sigma ^2}\left({\bf{P}},{{\bf{g}}_m},{{\bf{a}}_m}\right)}}\right)\;\;\forall \left( {m,l} \right)\;\;{\rm{s.}}{\rm{t.}}\;\;{a_{ml}} \ne 0\bigg\},
\end{align}
where ${\sigma ^2}\left({\bf{P}},{{\bf{g}}_m},{{\bf{a}}_m}\right)$ refers to the effective noise at the $m$th AP and ${\bf{P}} \buildrel \Delta \over = {\text{diag}}\left( {{P_1},{P_2}, \ldots ,{P_L}} \right)$ is the diagonal matrix with the power constraint for UEs.
In order to maximize the computation rate region, we need to find the optimal coefficient vector ${{{\bf{a}}_m}}$ and equalization factor $b_m$.

According to \cite[Lemma 2]{nazer2016expanding}, the equalization factor $b_m$ that minimizes the effective noise variance from (\ref{eff_noi}) is the MMSE projection.
Then, we have
\begin{equation}
{b_m} = {\bf{g}}_m^H {{\bf{P}}}{{\bf{a}}_m}{\left( {1 + {\bf{g}}_m^H{{\bf{P}}}{{\bf{g}}_m}} \right)^{ - 1}}.
\end{equation}
Hence, the effective noise is given by
\begin{align}
{\sigma ^2}\left( {{\bf{P}},{{\bf{g}}_m},{{\bf{a}}_m}} \right) & \buildrel \Delta \over = \frac{1}{n}{\mathbb{E}}\left\{ {{{\left\| {{{\bf{X}}^T}\left( {{b_m}{{\bf{g}}_m} - {{\bf{a}}_m}} \right) + {b_m}{{\bf{z}}_m}} \right\|}^2}} \right\}\notag\\
&= {\bf{a}}_m^H{\left( {{{{{\bf{P}}}}^{ - 1}} + {{\bf{g}}_m}{\bf{g}}_m^H} \right)^{ - 1}}{{\bf{a}}_m},
\end{align}
where ${\bf{X}} = {\left[ {{{\bf{x}}_1},{{\bf{x}}_2}, \ldots ,{{\bf{x}}_L}} \right]^T}$ represents the codeword matrix.
For the $m$th AP, the aim is to find its optimal coefficient vector that maximizes the computation rate region as
\begin{align}\label{optimize_a}
{{\bf{a}}_{m,\text{opt}}} &= \mathop {\arg \max }\limits_{{{\bf{a}}_m} \in \mathbb{Z}{{\left[ i \right]}^L}} {{\mathcal{R}}_{\text{ECF}}}\left( {{\bf{P}},{{\bf{g}}_m},{{\bf{a}}_m}} \right)\notag\\
&= \mathop {\arg \min }\limits_{{{\bf{a}}_m} \in \mathbb{Z}{{\left[ i \right]}^L}} {\sigma ^2}\left( {{\bf{P}}, {{\bf{g}}_m},{{\bf{a}}_m}} \right).
\end{align}
Since the channel coefficient between the $m$th AP and the $l$th UE is complex-valued, the received signal ${{\bf{y}}_m}$ can be divided into the real part and the imaginary part:
\begin{align}
&{\mathop{\rm Re}\nolimits} \left( {{{\bf{y}}_m}} \right)\!\! = \!\!\sum\limits_{l = 1}^L {\left( {{\mathop{\rm Re}\nolimits} \left( {{g_{ml}}} \right){\mathop{\rm Re}\nolimits} \left( {{{\bf{x}}_l}} \right)\!\! -\! \!{\mathop{\rm Im}\nolimits} \left( {{g_{ml}}} \right){\mathop{\rm Im}\nolimits} \left( {{{\bf{x}}_l}} \right)} \right)} \!\!+ \!\!{\mathop{\rm Re}\nolimits} \left( {{{\bf{z}}_m}} \right),\notag\\
&{\mathop{\rm Im}\nolimits} \left( {{{\bf{y}}_m}} \right)\!\!= \!\!\sum\limits_{l = 1}^L {\left( {{\mathop{\rm Im}\nolimits} \left( {{g_{ml}}} \right){\mathop{\rm Re}\nolimits} \left( {{{\bf{x}}_l}} \right)\!\! + \!\!{\mathop{\rm Re}\nolimits} \left( {{g_{ml}}} \right){\mathop{\rm Im}\nolimits} \left( {{{\bf{x}}_l}} \right)} \right)} \!\! + \!\!{\mathop{\rm Im}\nolimits} \left( {{{\bf{z}}_m}} \right).\notag
\end{align}
Therefore, we can transform the complex-valued network with $L$ UEs and $M$ APs into a real-valued network with $2 L$ UEs and $2 M$ APs. It is convenient to calculate the real and imaginary parts of the coefficient vector ${{\bf{a}}_m}$, respectively.
\footnote{Reducing the problem of developing coefficient algorithms for complex channels to an equivalent real-only channel is generally suboptimal.
Some solutions of finding the optimal solution in polynomial time over complex integer based lattices and complex channels were proposed in \cite{huang2016low}, however, they require a substantially higher complexity.
Therefore, the investigation of explicitly addresses the complex channel with low complexity is one of our future work.}
Without loss of generality, we only consider ${\mathop{\rm Re}\nolimits} \left( {{{\bf{a}}_m}} \right)$ for a given real-valued channel coefficient ${\mathop{\rm Re}\nolimits} \left( {{{\bf{g}}_m}} \right)$ in the following.

For each coefficient vector ${\mathop{\rm Re}\nolimits} \left( {{{\bf{g}}_m}} \right)$, we can find a signed permutation matrix $\textbf{S}$, which is unimodular and orthogonal such that ${\bf{S}}{\mathop{\rm Re}\nolimits} \left( {{{\bf{g}}_m}} \right)$ is nonnegative and its elements are in nondecreasing order \cite[Lemma 1]{zhou2014quadratic}.
Suppose ${\mathop{\rm Re}\nolimits} {\left( {{{\bf{a}}_m}} \right)_{\text{opt}}}$ is the optimal coefficient vector with the specifical power constraint $\textbf{P}$ and channel coefficient ${\mathop{\rm Re}\nolimits} \left( {{{\bf{g}}_m}} \right)$, we have $\mathcal{R}\left( {{\mathop{\rm Re}\nolimits} {\left( {{{\bf{g}}_m}} \right)},{\mathop{\rm Re}\nolimits} {\left( {{{\bf{a}}_m}} \right)_{\text{opt}}} }\right) = \mathcal{R}\left( {{\bf{S}}{\mathop{\rm Re}\nolimits} \left( {{{\bf{g}}_m}} \right),{\bf{S}}{\mathop{\rm Re}\nolimits} {\left( {{{\bf{a}}_m}} \right)_{\text{opt}}} }\right)$ \cite[Lemma 3]{zhou2014quadratic}.
Define $\overline {{\mathop{\rm Re}\nolimits} \left( {{{\bf{g}}_m}} \right)} $ as the nonnegative and non-decreasing-ordered vector, e.g., $\overline {{\mathop{\rm Re}\nolimits} \left( {{{\bf{g}}_m}} \right)}={\bf{S}}{\mathop{\rm Re}\nolimits} \left( {{{\bf{g}}_m}} \right)$. Therefore, we can recover the desired coefficient vector through ${\mathop{\rm Re}\nolimits} {\left( {{{\bf{a}}_m}} \right)_{\text{opt}}} = {{\bf{S}}^{ - 1}}{\overline {{\mathop{\rm Re}\nolimits} \left( {{{\bf{a}}_m}} \right)} _{\text{opt}}}$.

In the following, we concentrate on acquiring ${\overline {{\mathop{\rm Re}\nolimits} \left( {{{\bf{a}}_m}} \right)} _{\text{opt}}}$ for $\overline {{\mathop{\rm Re}\nolimits} \left( {{{\bf{g}}_m}} \right)} $ by relaxing the optimization problems stated in (\ref{optimize_a}) based on the quadratic programming (QP) method \cite{boyd2004convex}.
Recall that $\overline {{\mathop{\rm Re}\nolimits} \left( {{{\bf{a}}_m}} \right)} $ is in nondecreasing order, therefore, the maximum element should be ${\overline {{\mathop{\rm Re}\nolimits} \left( {{{\bf{a}}_m}} \right)} _L}$.
According to \cite{nazer2016expanding}, the searching space for ${\overline {{\mathop{\rm Re}\nolimits} \left( {{{\bf{a}}_m}} \right)} _L}$ can be restricted with
\begin{equation}\label{a_ml}
{\overline {{\mathop{\rm Re}\nolimits} \left( {{{\bf{a}}_m}} \right)} _L} \le {\lambda _{\max }}\left( {{\bf{I}} + \overline {{\mathop{\rm Re}\nolimits} \left( {{{\bf{g}}_m}} \right)}{{\bf{P}}}{{\overline {{\mathop{\rm Re}\nolimits} \left( {{{\bf{g}}_m}} \right)} }^T}} \right),
\end{equation}
where ${\lambda _{\max }}\left( {{\bf{I}} + \overline {{\mathop{\rm Re}\nolimits} \left( {{{\bf{g}}_m}} \right)} {{\bf{P}}}{{\overline {{\mathop{\rm Re}\nolimits} \left( {{{\bf{g}}_m}} \right)} }^T}} \right)$ denotes the maximum eigenvalue of $\left( {{\bf{I}} + \overline {{\mathop{\rm Re}\nolimits} \left( {{{\bf{g}}_m}} \right)} {{\bf{P}}}{{\overline {{\mathop{\rm Re}\nolimits} \left( {{{\bf{g}}_m}} \right)} }^T}} \right)$.
Then, the problem stated in (\ref{optimize_a}) can be rewritten as a series of QP problems
\begin{align}\label{QP}
\mathop {{\rm{minimize}}}\limits_{\overline {{\mathop{\rm Re}\nolimits} \left( {{{\bf{a}}_m}} \right)} }\qquad &{\overline {{\mathop{\rm Re}\nolimits} \left( {{{\bf{a}}_m}} \right)} ^T}{{\bf{G}}_m}\overline {{\mathop{\rm Re}\nolimits} \left( {{{\bf{a}}_m}} \right)} \notag\\
{\rm{subject\,to}}\qquad &\overline {{\mathop{\rm Re}\nolimits} \left( {{{\bf{a}}_m}} \right)}  \in {\mathbb{R}^L},\notag\\
&{\overline {{\mathop{\rm Re}\nolimits} \left( {{{\bf{a}}_m}} \right)} _L} = k,\;\;\;k = 1,2, \ldots ,K,
\end{align}
where ${{\bf{G}}_m} = {\left( {{\bf{P}} + \overline {{\mathop{\rm Re}\nolimits} \left( {{{\bf{g}}_m}} \right)}\, {\overline {{\mathop{\rm Re}\nolimits} \left( {{{\bf{g}}_m}} \right)} ^T}} \right)}^{-1}$ and $K = \left\lfloor {{\lambda _{\max }}\left( {{\bf{I}} + \overline {{\mathop{\rm Re}\nolimits} \left( {{{\bf{g}}_m}} \right)}{{\bf{P}}}{{\overline {{\mathop{\rm Re}\nolimits} \left( {{{\bf{g}}_m}} \right)} }^T}} \right)} \right\rfloor $.
Let $\overline {{\mathop{\rm Re}\nolimits} \left( {{{\bf{a}}_m}} \right)} _k^ + $ represent the solution to the problem of (\ref{QP}) with the constraint ${\overline {{\mathop{\rm Re}\nolimits} \left( {{{\bf{a}}_m}} \right)} _L} = k$.
$K$ solutions can be obtained by utilizing $\overline {{\mathop{\rm Re}\nolimits} \left( {{{\bf{a}}_m}} \right)} _k^ +  = k\overline {{\mathop{\rm Re}\nolimits} \left( {{{\bf{a}}_m}} \right)} _1^ + $.
With the Lagrange multiplier method \cite{boyd2004convex}, we have
\begin{equation}
\overline {{\mathop{\rm Re}\nolimits} \left( {{{\bf{a}}_m}} \right)} _1^ +  = \left[ {\bf{r}},1\right]^T,
\end{equation}
where ${\bf{r}} =  - {\left( {{{\bf{G}}_m}\left( {1:L - 1,1:L - 1} \right)} \right)^{ - 1}}{\bf{G}}\left( {1:L - 1,L} \right)$.
With the help of \cite[ALgorithm 1]{zhou2014quadratic}, $K$ real-valued solutions to the problem in (\ref{QP}), $\left\{ {\overline {{\mathop{\rm Re}\nolimits} \left( {{{\bf{a}}_m}} \right)} _k^ + } \right\}$, can be quantized to integer-valued $\left\{ {\overline {{\mathop{\rm Re}\nolimits} \left( {{{\bf{a}}_m}} \right)} _k^{{\mathop{\rm int}} }} \right\}$.
We select a sub-optimal coefficient vector ${\overline {{\mathop{\rm Re}\nolimits} \left( {{{\bf{a}}_m}} \right)} _{\text{opt}}}$ for $\overline {{\mathop{\rm Re}\nolimits} \left( {{{\bf{g}}_m}} \right)} $ with
\begin{equation}
{\overline {{\mathop{\rm Re}\nolimits} \left( {{{\bf{a}}_m}} \right)} _{\text{opt}}} = \arg \mathop {\min }\limits_{\overline {{\mathop{\rm Re}\nolimits} \left( {{{\bf{a}}_m}} \right)}  \in \left\{ {\overline {{\mathop{\rm Re}\nolimits} \left( {{{\bf{a}}_m}} \right)} _k^{{\mathop{\rm int}} }} \right\}} {\overline {{\mathop{\rm Re}\nolimits} \left( {{{\bf{a}}_m}} \right)} ^T}{{\bf{G}}_m}\overline {{\mathop{\rm Re}\nolimits} \left( {{{\bf{a}}_m}} \right)} .\notag
\end{equation}
Finally, the optimal coefficient vector ${\mathop{\rm Re}\nolimits} {\left( {{{\bf{a}}_m}} \right)_{\text{opt}}}$ correlated with the channel coefficient ${\mathop{\rm Re}\nolimits} \left( {{{\bf{g}}_m}} \right)$ is recovered with ${\mathop{\rm Re}\nolimits} {\left( {{{\bf{a}}_m}} \right)_{\text{opt}}} = {{\bf{S}}^{ - 1}}{\overline {{\mathop{\rm Re}\nolimits} \left( {{{\bf{a}}_m}} \right)} _{\text{opt}}}$.
Following a similar line of reasoning, the imaginary part of the coefficient vector can be derived.

\subsection{Parallel Computation}\label{Parallel Computation}
For parallel computation, the integer linear combinations of UEs' data are decoded independently.
On this basis, we first introduce the computation rate region.
Then, we provide a detailed description of the proposed power control algorithm which improves the achievable sum-rate.
For reducing the effective noise variance and computation complexity, we further propose an AP selection algorithm based on large-scale fading.

\subsubsection{Computation Rate Region}
Let us suppose that all APs have full channel state information.
To obtain the estimation of the integer combination with UEs' original data ${{{\bf{\hat u}}}_m}$, the $m$th AP multiplies the received signal by an equalization factor $b_m$ by the received signal to obtain the effective channel as
\begin{align}
&{\widetilde {\bf{y}}_m} = {b_m}{{\bf{y}}_m}= {b_m}{{\bf{X}}^T}{{\bf{g}}_m} + {b_m}{{\bf{z}}_m}\notag\\
&= {b_m}{{\bf{X}}^T}{{\bf{a}}_m} + \underbrace {{{\bf{X}}^T}\left( {{b_m}{{\bf{g}}_m} - {{\bf{a}}_m}} \right) + {b_m}{{\bf{z}}_m}}_{{\rm{effective\,noise}}}.
\end{align}
After choosing $b_m$ to be the minimum mean-square error (MMSE) coefficient adopted at the $m$th AP, the minimum effective noise variance for parallel computation is given by
\begin{equation}\label{sigema_para}
\sigma _{\text{para}}^2\left( {{\bf{P}},{{\bf{g}}_m},{{\bf{a}}_m}}\right) \buildrel \Delta \over = {\bf{a}}_m^H{\left( {{{{{\bf{P}}}}^{ - 1}} + {{\bf{g}}_m}{\bf{g}}_m^H} \right)^{ - 1}}{{\bf{a}}_m}.
\end{equation}
We denote ${\bf{A}}$ as the matrix of the coefficient vectors, ${\bf{A}} = \left[ {{{\bf{a}}_{1,}}{{\bf{a}}_2}, \ldots ,{{\bf{a}}_M}} \right]$.
Specifically, if the $m$th column of ${\bf{A}}$ is a null vector, the $m$th AP does not serve any UE; if the $l$th row of ${\bf{A}}$ is a zero vector, the $l$th UE is not served by any AP.
When we remove such columns and rows from ${\bf{A}}$, we obtain ${\bf{A}} \in {\mathbb{Z}}{\left[ i \right]^{L' \times M'}}$, where $L'$ and $M'$ refers to the number of effective UEs and APs.
Due to the array gain, the sum-rate increases along with the value of $M'$ increase.
However, there is a trade-off between the values of $L'$ and sum-rate performance, since the growth of effective UEs does not always lead to the increase in sum-rate \cite{ngo2017cell}.
According to the discussion of coefficient vector selection in Section III-A, the values of $M'$ and $L'$ are determined by the location of APs and UEs, therefore $M'$ and $L'$ can take the optimal value when the location of APs and UEs is optimal.
Define the rank of ${\bf{A}}$ by ${{M'}_{{\rm{rank}}}} \buildrel \Delta \over = {\rm{Rank}}\left( {\bf{A}} \right)$.
According to \cite{nazer2011compute}, all effective UEs' data can be recovered if ${{M'}_{{\rm{rank}}}} = L'$.
Therefore, we only need $L'$ integer linear combinations among the whole $M'$ combinations.
In other words, only $L'$ APs need to transmit signals to the CPU through fronthaul links.
The computation rate region for the parallel computation is given by
\begin{align}
&\!\!{\mathcal{R}_{{\rm{para}}}}\left( {{\bf{P}},{{\bf{g}}_m},{{\bf{a}}_m}} \right) \buildrel \Delta \over = \Bigg\{ {\left( {{R_1},{R_2}, \ldots ,{R_{L'}}} \right) \in \mathbb{R}_ + ^{L'}:}\notag\\
&\!\!{{R_l} \!\le\! {{\log }^ + }\!\left( \!{\frac{{{P_l}}}{{{\sigma _{\text{para}}^2}\left({{ {{\bf{P}},{{\bf{g}}_m},{{\bf{a}}_m}} }} \!\right)}}} \right)\;\;\forall \!\left( {m,l} \right)\;{\rm{s.}}{\rm{t.}}\;{a_{ml}} \ne 0} \Bigg\},
\end{align}
where $a_{ml}=0$ means the $m$th AP doesn't serve the $l$th UE.

\subsubsection{Power Optimization}
If ${a_{ml}} \ne 0$, the computed achievable rate for the $l$th UE at the $m$th AP is given as
\begin{equation}
{R'_{\left( {l,m} \right)}} = {\log ^ + }\left( {\frac{{{P_l}}}{{\sigma _{{\rm{para}}}^2\left( {{\bf{P}},{{\bf{g}}_m},{{\bf{a}}_m}} \right)}}} \right)
\end{equation}
However, when recovering the data of the $l$th UE, the codeword spacing for that data should tolerate the maximum effective noise variance across APs, whose linear combinations involve that data. Therefore, the actual achievable rate of the $l$th UE is
\begin{equation}
\!{R_l} \!=\! \mathop {\min }\limits_{{a_{ml}} \ne 0} \!{{R'}_{\left( {l,m} \right)}} \!=\! \mathop {\min }\limits_{{a_{ml}} \ne 0} {\log ^ + }\left( {\frac{{{P_l}}}{{\sigma _{{\rm{para}}}^2\left( {{\bf{P}},{{\bf{g}}_m},{{\bf{a}}_m}} \right)}}} \right).
\end{equation}
Hence, the achievable sum-rate of $L'$ UEs is
\begin{equation}\label{R}
\sum\limits_{l = 1}^{L'} {{R_l}}  = \mathop {\min }\limits_{{a_{ml}} \ne 0} \sum\limits_{l = 1}^{L'} {\left( {{{\log }^ + }\left( {\frac{{P_l}}{{\sigma _{{\rm{para}}}^2\left( {{{\bf{P}}},{{\bf{g}}_m},{{\bf{a}}_m}} \right)}}} \right)} \right)}.
\end{equation}
Recall that all UEs transmit with equal power in CF scheme.
For fairness, we compare the performance of CF and ECF with the constraint of equal total transmit power.
We aim at optimizing the power allocation to maximize the achievable sum-rate under the constraints on the total power consumption $P_t$. The optimization problem is formulated as follows:
\begin{align}
\mathop {{\rm{maximize}}}\limits_{\bf{P}} &\sum\limits_{l = 1}^{L'} {{R_l}} \notag\\
{\rm{subject\,to}}&\sum\nolimits_{l = 1}^{L'} {{P_l}}  = {P_t},\notag\\
&{P_l} \ge 0,l = 1,2, \ldots .L'.
\end{align}
UEs can share a total power budget which is the upper bound performance of each UE's power constraint as their total maximum allowable transmit power \cite{he2018uplink}.
Besides, (19) is handled at the CPU since the global information ${{\bf{a}}_1}, \cdots ,{{\bf{a}}_M}$ and ${{\bf{g}}_1}, \cdots ,{{\bf{g}}_M}$ are required.

As mentioned above, each AP decodes $\sum\nolimits_{l = 1}^L {{a_{ml}}{{\mathbf{x}}_l}}$ as one regular codeword due to the lattice algebraic structure.
All UEs served by the $m$th AP need to tolerate the same effective noise.
If the linear integer combinations $\sum\nolimits_{l = 1}^L {{a_{ml}}{{\mathbf{x}}_l}}$ can tolerate the effective noise variance ${{{\sigma _{{\rm{para}}}^2\left( {{{\bf{P}}},{{\bf{g}}_m},{{\bf{a}}_m}} \right)}}}$, then all UEs served by the $m$th AP, which means ${a_{ml}} \ne 0$ can be successfully recovered from the linear combination with integer coefficient vector ${{\bf{a}}_m}$.
In cell-free massive MIMO, we always emphasize a good quality-of-service for all users.
However, directly improving the achievable sum rate cannot achieve a good balance of quality-of-service for all users \cite{bjornson2017massive}.
Therefore, the goal of minimizing the maximum effective noise variance that can generally improve the achievable rate for most UEs is more suitable for our model.
In other words, we could minimize the maximum effective noise variance as
\begin{align}\label{para_op}
\mathop {{\rm{minimize}} }\limits_{{{\bf{P}}}} \mathop {\max }\limits_{m = 1, \ldots ,M'}& \left\{ {{{\sigma _{{\rm{para}}}^2\left( {{{\bf{P}}},{{\bf{g}}_m},{{\bf{a}}_m}} \right)}}} \right\}\notag\\
{\rm{subject\,to}}&\sum\nolimits_{l = 1}^{L'} {{P_l}}  = P_t,\notag\\
&{P_l} \ge 0,\;\;\;l = 1,2, \ldots .L'.
\end{align}
According to (\ref{sigema_para}), (\ref{para_op}) is equivalent to
\begin{align}\label{para_sigema_op}
\mathop {{\rm{minimize}} }\limits_{\bf{P}} \mathop {\max }\limits_{m = 1, \ldots ,M'} &\left\{ {{\bf{a}}_m^H{\left( {{{{{\bf{P}}}}^{ - 1}} + {{\bf{g}}_m}{\bf{g}}_m^H} \right)^{ - 1}}{{\bf{a}}_m}} \right\}\notag\\
{\rm{subject\,to}}&\sum\nolimits_{l = 1}^{L'} {{P_l}}  = {P_t},\notag\\
&{P_l} \ge 0,\;\;\;l = 1, \ldots ,L'.
\end{align}
According to \cite[Lemma B. 4]{bjornson2017massive}, the matrix inversion can be equivalently represented by
\begin{equation}
{\left( {{{ {{{\bf{P}}}}}^{ - 1}} + {{\bf{g}}_m}{\bf{g}}_m^H} \right)^{ - 1}} = {{\bf{P}}} - \frac{1}{{1 + {\bf{g}}_m^H{{\bf{P}}}{{\bf{g}}_m}}}{{\bf{P}}}{{\bf{g}}_m}{\bf{g}}_m^H{{\bf{P}}}.
\end{equation}
Therefore, the effective noise variance for the $m$th AP is
\begin{equation}\label{para_sigema}
\sigma _{{\rm{para}}}^2\left( {{\bf{P}},{{\bf{a}}_m},{{\bf{g}}_m}} \right) = {\bf{a}}_m^H{\bf{P}}{{\bf{a}}_m} - \frac{{{\bf{a}}_m^H{\bf{P}}{{\bf{g}}_m}{\bf{g}}_m^H{\bf{P}}{{\bf{a}}_m}}}{{1 + {\bf{g}}_m^H{\bf{P}}{{\bf{g}}_m}}}.
\end{equation}
Aa a result, (\ref{para_sigema_op}) can be rewritten as
\begin{align}\label{NP}
\mathop {{\rm{minimize}}}\limits_{\bf{P}} \mathop {\max }\limits_{m = 1, \ldots ,M'}& {\left\{ {{\bf{a}}_m^H{\bf{P}}{{\bf{a}}_m} - \frac{{{\bf{a}}_m^H{\bf{P}}{{\bf{g}}_m}{\bf{g}}_m^H{\bf{P}}{{\bf{a}}_m}}}{{1 + {\bf{g}}_m^H{\bf{P}}{{\bf{g}}_m}}}} \right\}}\notag\\
{\rm{subject\,to }}&\sum\limits_{l = 1}^{L'} {{P_l}}  = {P_t},\notag\\
&{P_l} \ge 0,\;\;\;l = 1, \ldots ,L'.
\end{align}
However, (\ref{NP}) is NP-hard. To tackle this challenge, we first introduce three auxiliary variables. On this basis, we can build the following optimization problem by introducing three auxiliary variables $r$, $s$, and $t$:
\begin{equation}\label{052801}
\left\{
\begin{aligned}
\mathop {\min }\limits_{{\bf{P}}, r, s}\;\;\;&t\\
{\rm{s}}{\rm{.t}}{\rm{.}}\;\;\;
&t \ge {r} - {s},\;\;\;\forall m,\\
&{r} \ge {\bf{a}}_m^H{\bf{P}}{{\bf{a}}_m},\;\;\;\forall m,\\
&{s} \le \frac{{{{\left\| {{\bf{a}}_m^H{\bf{P}}{{\bf{g}}_m}} \right\|}^2}}}{{1 + {\bf{g}}_m^H{\bf{P}}{{\bf{g}}_m}}},\;\;\;\forall m,\\
&\sum\limits_{l = 1}^{L'} {{P_l}}  = {P_t},\\
&{P_l} \ge 0,\;\;\;\forall l.
\end{aligned}
\right.
\end{equation}
In particular, variables $r$ and $s$ have limited searching space, respectively. For a given value of $r$, the variable $s$ should be smaller than $r$.
Therefore, we employ a two-dimension of brute force search on these two scalars.
For fixed $r$ and $s$, the optimization problem in (\ref{052801}) can be rewritten as a feasibility problem
\begin{equation}\label{052802}
\left\{
\begin{aligned}
&{\bf{a}}_m^H{\bf{P}}{{\bf{a}}_m} \le r,\;\;\;\forall m,\\
&\left( {1/2} \right){{\bf{p}}^T}{\bf{Jp}} - s{{\bf{v}}}{\bf{p}} - {{s}} \ge {{0}},\;\;\;\forall m,\\
&\sum\limits_{l = 1}^{L'} {{P_l}}  = {P_t},\\
&{P_l} \ge 0,\;\;\;\forall l,
\end{aligned}
\right.
\end{equation}
where ${\bf{p}} = {\left[ {{P_1}, \ldots ,{P_{L'}}} \right]^T}$, ${\bf{v}} = {\left[ {{{\left| {{g_{m1}}} \right|}^2}, \ldots ,{{\left| {{g_{mL'}}} \right|}^2}} \right]^T}$. ${\bf{J}}$ is a $L' \times L'$ matrix whose ${\left( {{l_1},{l_2}} \right)}$th element is given as ${{\bf{J}}_{\left( {{l_1},{l_2}} \right)}} = 2{\left| {{a_{m{l_1}}}} \right|\left| {{g_{m{l_1}}}} \right|\left| {{a_{m{l_2}}}} \right|\left| {{g_{m{l_2}}}} \right|}$.
Clearly, ${\bf{J}}$ is a positive semi-definite matrix.
Consequently, (\ref{052802}) can be solved efficiently by performing a brute force search on two scalars. In each step, a feasibility problem needs to be solved.
Since transforming a nonconvex problem into its equivalent convex form is quite difficult if not possible, an off-the-shelf optimization solver, e.g. fmincon in Matlab, is adopted to obtain a suboptimal solution.
Besides, simulations show that solving a nonconvex problem also brings obvious performance improvement, the computation cost is tolerable.
Besides, actually (\ref{NP}) is equal to (\ref{052801}) only if the two terms in the objective of (\ref{NP}) are independent.
However, at the optimal point, we only care about minimize the final maximum term.
More specifically, Algorithm \ref{para_algorithm} can solve (\ref{052801}).
The parameters in Step 1, e.g., ${r_{\min }}$ and ${r_{\max }}$, can be determined by solving another two feasibility problems:
\begin{equation}
\begin{cases}
\begin{array}{l}
\mathop {\max }\limits_{\bf{P}}\;\;\;{\bf{a}}_m^H{\bf{P}}{{\bf{a}}_m}\\
{\rm{s}}{\rm{.t}}{\rm{.}}\;\;\;\sum\limits_{l = 1}^{L'} {{P_l}}  = {P_t},\\
\;\;\;\;\;\;\;{P_l} \ge 0,\;\;\;\forall l,
\end{array}
\end{cases}
\quad
\begin{cases}
\begin{array}{l}
\mathop {\min }\limits_{\bf{P}}\;\;\;{\bf{a}}_m^H{\bf{P}}{{\bf{a}}_m}\\
{\rm{s}}{\rm{.t}}{\rm{.}}\;\;\;\sum\limits_{l = 1}^{L'} {{P_l}}  = {P_t},\\
\;\;\;\;\;\;{P_l} \ge 0,\;\;\;\forall l,
\end{array}
\end{cases}
\end{equation}
respectively. When the search is completed, all APs achieve the same minimal effective noise variance $t$.
The corresponding values of $r$ and $s$ can be denoted as $r_{\text{opt}}$ and $s_{\text{opt}}$.
Finally, utilizing (\ref{R}) and (\ref{para_sigema}), the achievable sum-rate can be obtained.
In Algorithm 1, there are at most
\begin{equation}
  \left\lfloor \!{\frac{{\left(\! {2{r_{\max }} \!-\! {r_{sl}}\!\left\lfloor {\frac{{{r_{\max }} \!-\! {r_{\min }}}}{{{r_{sl}}}} \!- \!1} \right\rfloor } \!\right)\!\left\lfloor {\frac{{{r_{\max }} \!-\! {r_{\min }}}}{{{r_{sl}}}}} \right\rfloor }}\!{{2{s_{sl}}}}} \right\rfloor \!\!\left\lfloor \!{\frac{{{r_{\max }} \!-\! {r_{\min }}}}{{{r_{sl}}}}} \!\right\rfloor
\end{equation}
feasibility problems that need to be solved, where ${s_{sl}}$ and ${r_{sl}}$ refer to the step size for searching $r$ and $s$, respectively.

\begin{algorithm}[t]
\caption{Brute Force Search on Two Scalars for Solving Problem (\ref{052801})}
\label{para_algorithm}
\begin{algorithmic}[1]
\State \emph{Initialization:} Define the range of the values of $r$ by ${r_{\min }}$ and ${r_{\max }}$.
        Choose step size for $r$ and $s$ as $r_{sl}$ and $s_{sl}$, respectively. Set $t=\infty$ and $r = {r_{\max }}$.
\State Set $s = r- s_{sl}$. If $s \ge 0$, solving the feasibility program in (\ref{052802}),else go to Step Step 4.
\State If (\ref{052802}) is feasible, set $t = \min \left( {t,r - s} \right)$, else back to Step Step 2.
\State Update $r$ with $r = r - r_{sl}$. Stop if $r \le {r_{\min }}$.
\end{algorithmic}
\end{algorithm}

\subsubsection{AP Selection}\label{AP selection}
As mentioned above, recovering $L'$ UEs' original data only requires $L'$ integer linear combinations.
Therefore, we propose a low-complexity AP selection algorithm for two purposes.
First, only $L'$ effective noise variance participant in the brute force search leads to a reduction in computational complexity.
Second, the noise tolerance on UEs' data can be relaxed, which contributes to the improvement of the achievable sum-rate.
Recall that (\ref{a_ml}) restricts the maximum value in the coefficient vector ${{{\bf{a}}_m}}$ and the search space is generally small.
Hence, for different APs, it is the difference on the second term in (\ref{para_sigema}) that leads to a significant deviation on the effective noise variance.

Note that the average channel gain is -70 dB while the noise power is -130 dBW.
Therefore, the denominator of that term is close to 1 and is several orders of magnitude smaller than the numerator.
Consequently, the main factor that affects the effective noise variance across different APs is the sum of all elements in ${\bf{J}}$.
In other words, decoding the estimations ${{{\bf{\hat u}}}_m}$ of APs with a high sum value of ${\bf{J}}$ will obtain a higher achievable sum-rate.
Therefore, we prefer selecting APs with a high sum value of ${\bf{J}}$.
Furthermore, according to (\ref{g_ml}), $g_{ml}$ is a function of $\beta_{ml}$.
Then, we propose an algorithm for AP selection based on large-scale fading coefficient ${\beta _{ml}}$, which generally stays constant for several coherence intervals.

We construct matrix ${\bf{J}}$ for each AP by replacing the channel coefficient $g_{ml}$ with ${\beta _{ml}}$.
For each ${\bf{J}}$, we first sum all the elements and sort the sum values in ascending order.
Then, we apply the greedy AP selection for message recovery stated in \cite[Algorithm 1]{huang2017compute}.
Compared with the AP selection in \cite{huang2017compute}, we sort the APs by firstly replacing the channel coefficient $g_{ml}$ with $\beta_{ml}$, and then calculating the sum value of all the elements in the matrix $J$, which is independent of the power allocation.
According to \cite{huang2017compute}, we check the columns of ${\bf{A}}$ one by one, where ${\bf{A}} = {\left[ {{\bf{a}}_1 \ldots ,{\bf{a}}_M} \right]}$, until rank requirement satisfied, therefore the computational complexity of the proposed AP selection method is no more than ${\cal O}\left( {M' + M'{{\log }_2}\left( {M'} \right) + M'{{\left( {M' - 1} \right)}^3}} \right)$.

\subsection{Successive Computation}\label{Successive Computation}
It is beneficial to remove the codewords which have been decoded successfully from the channel observation.
In that case, subsequent decoding stages will encounter less interference.
This well-known technique is referred to as successive interference cancellation (SIC).
In ECF framework, we apply an analog of that for cell-free massive MIMO, which is named successive computation and can be viewed as the combination of ECF and successive interference cancellation.
Compared with parallel computation, successive computation reduces the effective noise variance and the number of users that need to tolerate that effective noise in each decoding step \cite{nazer2016expanding}.
Hence, it can further improve the system performance.
The SIC technique also applied in \cite{al2020multiple}, which proposes a hybrid deep reinforcement learning (DRL) model to design the IUI-aware receive diversity combining scheme.
Compared with \cite{al2020multiple}, our successive computation scheme benefits from applying the nested lattice coding strategy which can effectively reduce the fronthaul load.
In this subsection, the expression of the computation rate region for successive computation is introduced firstly.
Since the decoding order of integer linear combinations ${{{\bf{\tilde u}}}_m}$ and UEs both have a significant impact on the performance of successive computation, we present different methods to find the sub-optimal decoding orders with power control.

\subsubsection{Computation Rate Region}
In successive computation, AP $m$ sends the received signal ${\bf{y}}_m$ and the integer linear combinations of codewords ${\bf{a}}_m^T{\bf{X}}$ to the CPU, instead of decoding the received signal ${{\bf{y}}_m}$ into the combinations of UEs' original data ${{{\bf{\hat u}}}_m}$
\footnote{When applying with successive computation, the interference cancellation procedure takes place at the decoder equipped at the CPU.
Note that the signaling exchanges occurred per coherent interval and the fronthaul load are $2M'n $ and $4M'n $ with parallel computation and successive computation, respectively.
Besides, the data of UEs are conveyed from the CPU to the APs through fronthaul links and then distributed to the UEs with precoding.}.
Define ${{\bf{A}}_{m - 1}} \buildrel \Delta \over = {\left[ {{{\bf{a}}_1}, \ldots ,{{\bf{a}}_{m - 1}}} \right]^T}$, the CPU applies equalization factor $b_m$ and vector ${\bf{c}}$ to the $m$th combination as
\begin{align}\label{33}
&{{{\bf{\tilde y}}}_m} = {b_m}{{\bf{y}}_m} + {{\bf{X}}^T}{{\bf{A}}_{m - 1}}{\bf{c}}\notag\\
&= {{\bf{X}}^T}{{\bf{a}}_m} + \underbrace {{{\bf{X}}^T}\left( {{b_m}{{\bf{y}}_m} + {{\bf{A}}_{m - 1}}{\bf{c}} - {{\bf{a}}_m}} \right) + {b_m}{{\bf{z}}_m}}_{{\rm{effective\;noise}}}.
\end{align}
where ${\bf{c}} = {\left[ {{c_1}, \ldots ,{c_{m - 1}}} \right]^T}$. (\ref{33}) shows that the decoded linear combinations $\left\{{{{\bf{y}}}_1}, \ldots ,{{{\bf{y}}}_{m-1}}\right\}$ can be used as side information for reducing the effective noise experienced in the latter decoding stages for other UEs.

After choosing $b_m$ and $\textbf{c}$ to be the MMSE projection scalar and vector, the minimum effective noise variance with transmit power matrix $\textbf{P}$ is given by
\begin{equation}\label{sigema_succ}
\sigma _{{\rm{succ}}}^2\left( {{{\bf{g}}_m},{{\bf{a}}_m},\left. {\bf{P}} \right|{{\bf{A}}_{m - 1}}} \right) \buildrel \Delta \over = {\bf{a}}_m^H{{\bf{F}}^T}{{\bf{N}}_{m - 1}}{\bf{F}}{{\bf{a}}_m},
\end{equation}
where
\begin{align}
{{\bf{F}}^T}{\bf{F}}  = {\left( {{{\bf{P}}^{ - 1}} + {{\bf{g}}_m}{\bf{g}}_m^H} \right)^{ - 1}},
\end{align}
and
\begin{align}
{{\bf{N}}_{m - 1}}  \!= \!{\bf{I}} \!-\! {\bf{FA}}_{m - 1}^T{\left( {{{\bf{A}}_{m - 1}}{{\bf{F}}^T}{\bf{FA}}_{m - 1}^T} \right)^{ - 1}}{{\bf{A}}_{m - 1}}{{\bf{F}}^T}.
\end{align}
Then, the computation rate region for successive computation is given as
\begin{align}\label{R_succ}
&{\mathcal{R}_{{\rm{succ}}}}\left( {{\bf{P}},{{\bf{g}}_m},\left. {{{\bf{a}}_m}} \right|{{\bf{A}}_{m - 1}}} \right) \buildrel \Delta \over = \Bigg\{ {\left( {{R_1}, \ldots ,{R_{L'}}} \right) \in R_ + ^{L}:}\notag\\
&{{R_l} \!\le \!{{\log }^ + }\!\left( \!{\frac{{{P_l}}}{{\sigma _{{\rm{succ}}}^2\left( {{\bf{P}},{{\bf{g}}_m},\left. {{{\bf{a}}_m}} \right|{{\bf{A}}_{m - 1}}} \right)}}} \!\right)\forall \!\left( {m,l} \right)\;{\text{s.t.}}\;{a_{ml}} \!\ne\! 0} \!\Bigg\}.
\end{align}

\subsubsection{Searching Decoding Order for Combinations}
Among $M'$ APs, we first select the effective UEs and APs that participate in the uplink transmission.
Then, we try to find the candidate integer combinations with small effective noise utilizing the AP selection algorithm.
The detailed procedure has been introduced in Section \ref{AP selection}.
According to (\ref{R_succ}), the effective noise variance is related to the decoding order of integer linear combinations.
Therefore, we propose an efficient method for determining the side information matrix ${{\bf{A}}_{m - 1}}$ with power control.

The main idea of successive computation is to utilize the side information and the decoded integer linear combinations to reduce the effective noise.
Note that the integer linear combination decoded firstly does not have any side information to exploit.
Hence, the effective noise expression for the first decoding step is similar to that of parallel computation, which is given by
\begin{align}\label{SUCC1}
&\sigma _{{\rm{succ}},\delta \left( 1 \right)}^2\left( {{\bf{P}},{{\bf{g}}_{\delta \left( 1 \right)}},{{\bf{a}}_{\delta \left( 1 \right)}}} \right) \notag\\
&= {\bf{a}}_{\delta \left( 1 \right)}^H{\left( {{{\bf{P}}^{ - 1}} + {{\bf{g}}_{\delta \left( 1 \right)}}{\bf{g}}_{\delta \left( 1 \right)}^H} \right)^{ - 1}}{{\bf{a}}_{\delta \left( 1 \right)}}.
\end{align}
where $\delta \left( m \right)$ denotes the decoding order.
As the remaining combinations can reduce their effective noise with the help of side information, we can select the integer linear combination which has the minimum effective noise to decode firstly.
To determine ${\bf{a}}_{\delta \left( 1 \right)}^T{\bf{X}}$, we solve the following optimization problem to obtain its local sub-optimal power allocation.
As the problem (\ref{SUCC_001}) is non-convex and translate it into non-convex is quite difficult, an off-the-shelf optimization solver, e.g. \textit{fmincon} in Matlab, is adopted to obtain a suboptimal solution.
Note that although the obtained solution is suboptimal, the performance of the proposed framework is still superior compared with CF and MRC, which will be verified in the simulation section.
Such that, for each ${\bf{a}}_m^T{\bf{X}}$, $m= 1, \ldots ,L'$, we have
\begin{align}\label{SUCC_001}
\mathop {\min }\limits_{\bf{P}}\,\, &\sigma _{{\rm{succ}},\delta \left( 1 \right)}^2\left( {{\bf{P}},{{\bf{g}}_m},{{\bf{a}}_m}} \right)\notag\\
{\rm{s}}{\rm{.t}}{\rm{.}}\,\,&\sum\nolimits_{l = 1}^{L'} {{P_l}}  = {P_t},\notag\\
&{P_l} \ge 0,l = 1, \ldots ,L'.
\end{align}
Then, we calculate the effective noise $\sigma _{{\rm{succ}},\delta \left( 1 \right)}^2$ for each ${\bf{a}}_m^T{\bf{X}}$ with its own power allocation matrix and select the combination which has the minimal effective noise as ${\bf{a}}_{\delta \left( 1 \right)}^T{\bf{X}}$.

After determining ${\bf{a}}_{\delta \left( 1 \right)}^T{\bf{X}}$, we begin to determine the remaining decoding order.
For the $m$th step, we determine ${{\bf{a}}^{T}_{\delta \left( m \right)}}{\bf{X}}$ depending on the effective noise and the rank of the side information matrix.
More specifically, we first calculate the effective noise variance for each of the remaining integer linear combinations according to (\ref{sigema_succ}) and sort them in ascending order. Then, in line with the order, in each turn add the corresponding coefficient vector to the side information matrix ${{\bf{A}}_{m - 1}}$, which is known from step 1 through $m-1$, to form ${{\bf{A}}_{m}} = \left[ {{{\bf{A}}_{m - 1}};{\bf{a}}_{m'}^T} \right]$, $m' = 1, \ldots ,M' - m + 1$.
Finally, we select the integer linear combination which meets the constraint
\begin{equation}\label{RANK}
{\rm{Rank}}\left( {{{\bf{A}}_m}} \right) = m,
\end{equation}
to update the side information matrix.
The procedure for finding ${{\bf{a}}_{\delta \left( m \right)}^T}{\bf{X}}$ terminates when all integer linear combinations find themselves decoding orders.
The detailed procedure for searching the decoding order of combinations in successive computation is summarized in Algorithm \ref{SUCC_algorithm}.
To determine ${\bf{a}}_{\delta \left( 1 \right)}^T{\bf{X}}$, we need to solve the optimization problem in (\ref{SUCC_001}) for $L'$ times.
Then, for searching the decoding order for the left $L'-1$ combinations in terms of the number of complex multiplications is
\begin{align}
&\sum\limits_{m = 2}^{L'} {\frac{{\left( {L' - m + 2} \right)\left( {L' - m + 1} \right)}}{2}}{\left[ {\frac{{{{\left( {L'} \right)}^3} - L'}}{3}}+ {{\left( {m - 1} \right)}^3}L' \right.}\notag\\
&{\left. {+ \frac{{{{\left( {m - 1} \right)}^2}L' \!\!+\!\! \left( {m - 1} \right)L'}}{2} \!\! +\!\! \left( {m -1} \right){{\left( {L'} \right)}^2} \!\!+\!\! 2{{\left( {L'} \right)}^2}} \right]}.
\end{align}

\begin{algorithm}[t]
\caption{Searching Decoding Order for Combinations in Successive Computation}
\label{SUCC_algorithm}
\begin{algorithmic}[1]
\State $\textbf{Input}$: $L'$ candidate integer linear combinations and the corresponding coefficient vectors which can be obtained through AP selection.
\State \emph{Initialization:} $m=2$, $n=1$.
\State Solve the optimization problem (\ref{SUCC_001}) for all $L'$ combinations and calculate the effective noise variance for each combination using (\ref{SUCC1}).
Select the combination which has the minimal effective noise as ${\bf{a}}_{\delta \left( 1 \right)}^T{\bf{X}}$ and remove the corresponding coefficient vector from the candidate set.
The side information is obtained with ${{\bf{A}}_{1}} = \left[ {\bf{a}}_{\delta \left( 1 \right)}^T \right]$.
The power constraint matrix ${\bf{P}}$ is determined with the power allocation for ${\bf{a}}_{\delta \left( 1 \right)}^T{\bf{X}}$.
\State Calculate the effective noise variance for each of the remaining combinations based on ${{\bf{A}}_{m-1}}$ using (\ref{sigema_succ}) and sort them in ascending order.
Update ${{\bf{A}}_{m}}$ with ${{\bf{A}}_{m-1}} = \left[ {{\bf{A}}_{m-1}}; {{\bf{a}}_n^T} \right]$. If ${\rm{Rank}}\left( {{{\bf{A}}_m}} \right) = m$, remove ${{\bf{a}}_n}$ form the candidate set to update the side information matrix and set $m:=m+1, n=1$, else $n:=n+1$. \label{SUCC2}
\State Stop if ${\rm{Rank}}\left( {{{\bf{A}}_L'}} \right) = L'$, otherwise back to Step \ref{SUCC2}.
\State $\textbf{Output}$: ${{{\bf{A}}_L'}}$.
\end{algorithmic}
\end{algorithm}

\subsubsection{Searching Decoding Order for UEs}
In successive computation, the effective noise of UEs whose data is decoded in the latter decoding stages can be reduced.
At the $m$th decoding step, it is possible to use the side information matrix ${{\bf{A}}_{m - 1}}$ to reduce some known individual codewords and remove them from the integer linear information ${\bf{a}}_{\delta \left( m \right)}^T{\bf{X}}$ without changing the effective noise variance.
In particular, if the $l$th UE's data has been recovered at the $m$th step, the data only need to tolerate the maximum effective noise among $\left\{ { {\bf{a}}_{\delta \left( 1 \right)}^T{\bf{X}}, \ldots ,{\bf{a}}_{\delta \left( m \right)}^T{\bf{X}} } \right\}$.
Therefore, the decoding order of UEs also has an effect on the achievable sum-rate. Searching the decoding order has been studied in some works \cite{mesbah2009joint}, \cite{zhou2011joint} with SIC.
However, our problem is generally intractable and the use of convex optimization for obtaining the optimal solution is not possible.
Therefore, we propose several methods to determine the decoding order of UEs and select the best one as a suboptimal solution.

\paragraph{Received-Power-Based Algorithm}
In general successive interference cancellation, the decoding order of UEs is determined by their received power.
We calculate the received power of the $l$th UE with respect to the $m$th integer linear combination with
${P_{r,\left( {l,m} \right)}} = {P_l}{\left\| {{g_{ml}}} \right\|^2}$.
Then, we can obtain the achievable rate for each UE with ${\bf{a}}_{\delta \left( m \right)}^T{\bf{X}}$ with
\begin{equation}\label{R_ml}
{R_{{\delta \left( m \right)},l}} = \frac{{{P_l}}}{{\sigma _{{\rm{succ}},\delta \left( m \right)}^2\left( {{\bf{P}},{{\bf{g}}_{\delta \left( m \right)}},\left. {{{\bf{a}}_{\delta \left( m \right)}}} \right|{{\bf{A}}_{m - 1}}} \right)}},\;{a_{{\delta \left( m \right)},l}} \ne 0,
\end{equation}
and sort them in descending order based on the received power.
The signal from UE whose rate with respect to the combination is in the first place of the order is decoded at the ${\delta \left( m \right)}$ step.

\paragraph{Channel-Coefficient-Based Algorithm}
As stated in Section \ref{AP selection}, a better channel condition leads to the less effective noise variance.
Therefore, UEs with good channel condition contributes to the small effective noise of selected APs.
These UEs can be decoded first to relax their effective noise tolerance and then have a large achievable rate.
For the $l$th UE, let us define ${{\bf{g}}_l}$ as the channel coefficients with $L'$ integer linear combinations ${{\bf{g}}_l} = {\left[ {{g_{1l}}, \ldots ,{g_{L'l}}} \right]}^T$.
We calculate the 2-norm of ${{\bf{g}}_l}$ for all UEs and sort them in descending order.

\paragraph{Hungarian Algorithm}
With the effective noise variance for $L'$ integer linear combinations and the power allocation for $L'$ UEs, we can find the assignment for each UE with $P_l$.
It has been shown in \cite{patel2017scheduling} that the Hungarian algorithm may be the best solution to the combinational optimization problem.
Therefore, we first construct a $L' \times L'$ matrix ${\bf{C}}$, whose element in the $l$th row and $m$th column represents the achievable rate of the $l$th UE with the $m$th integer linear combination from (\ref{R_ml}), such as $C_{ml}=R_{ml}$.
The conventional Hungarian algorithm aims to find $L'$ element which are set in different rows and columns of ${\bf{C}}$.
The sum of these $L'$ element is minimum.
However, we need to obtain the maximum value of the achievable sum-rate.
Therefore, we first find the maximum value ${C_{\max }}$ in ${\bf{C}}$ and replace each element with ${C_{ml}} = {C_{ml}} - {C_{\max }}$. The detailed procedure of the Hungarian algorithm is summarized in Algorithm \ref{hungarian_algorithm}.

\begin{algorithm}[t]
\caption{Hungarian Algorithm for Finding Optimal Decoding Order of UEs}
\label{hungarian_algorithm}
\begin{algorithmic}[1]
\State Perform row operations on ${\bf{C}}$.
The minimum element of each row is selected and is subtracted from each element in that row.\label{row}
\State Repeat the procedure stated in Step \ref{row} for all columns.
\State Count the minimum number of rows and columns that cover all zeros. Test the optimality.
If the number of counted lines is equal to $L'$, such as ${N_l} = L'$, stop the procedure.\label{3}
\State Find the minimum value that is not covered in lines and add that to intersection points.
Subtract that minimum value from elements that are not covered by counted lines.\label{4}
\State Repeat Step \ref{3} for checking the optimality condition. If ${N_l} \le L'$, repeat Step \ref{4}.
\end{algorithmic}
\end{algorithm}

To determine which UE's data should be recovered at the $m$th decoding step, we need to compute $L' + 1 - m$ UEs' achievable rate.
Therefore, the computational complexity for the received-power-based algorithm and the channel-coefficient-based algorithm is ${\cal O}\left( {\frac{{\left( {L' + 1} \right)L'}}{2}} \right)$.
Besides, the complexity of the Hungarian algorithm is ${\cal O}\left( {{{\left( {L'} \right)}^3}} \right)$ \cite{xu2007study}.
After determining the decoding order of integer linear combinations and UEs, the achievable sum-rate can be obtained.
Using (\ref{R_ml}) for calculating the achievable rate for $l$th UE whose data is recovered with ${\bf{a}}_{\delta \left( m \right)}^T{\bf{X}}$, and the sum achievable rate is given as ${R_{\text{sum}}} = \sum\nolimits_{l = 1}^{L'} {{R_{ml}}} $.

\section{Numerical Results}\label{numerical_results}
\subsection{Parameters Setup}
We adopt the similar parameters setting in \cite{ngo2017cell} as the basis to establish our simulation system model. More specifically, all UEs and APs are randomly located within a square of 1 $\times $ 1 km.
In each simulation setup, the APs and UEs are uniformly distributed at random locations within the simulation area.
The square is wrapped around at the edges to avoid boundary effects.
Hata-COST231 model is employed to characterize the large-scale propagation.

\subsection{Results and Discussion}
\subsubsection{Parallel Computation}
\begin{figure}[t!]
\centering
\includegraphics[width=3.5in]{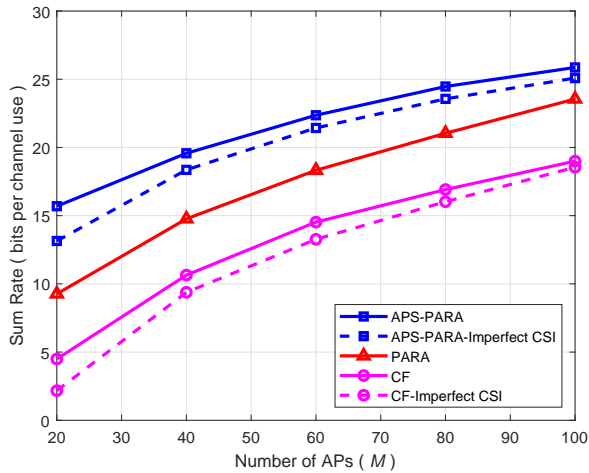}
\caption{Achievable sum-rate for CF, PARA, and APS-PARA schemes with $L=10$ and ${P_{t}=200}$ mW.}
\label{fig:1}
\end{figure}

\begin{figure}[t!]
\centering
\includegraphics[width=3.5in]{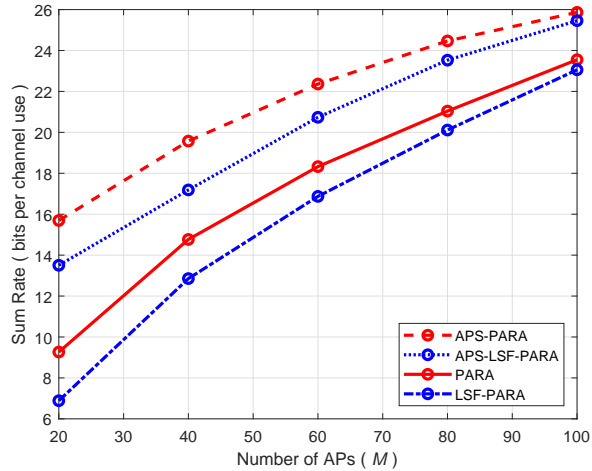}
\caption{Achievable sum-rate for PARA, APS-PARA, LSF-PARA, and APS-LSF-PARA schemes with $L=10$ and ${P_{t}=200}$ mW.}
\label{fig:2}
\end{figure}

First, we evaluate the performance of the proposed parallel computation (PARA) scheme in terms of the achievable sum-rate with power control.
The APS-PARA scheme refers to the PARA with AP selection.
Fig. \ref{fig:1} shows the achievable sum-rate obtained via CF, PARA, and APS-PARA schemes versus the number of APs with $L=10$ and ${P_{t}=200}$ mW.
Owing to the array again, the system performance of all considered schemes increases as the number of APs $M$ increasing.
Moreover, the PARA scheme with the proposed power control method outperforms the conventional CF scheme.
For example, compared with the CF scheme, both PARA and APS-PARA schemes improve the achievable sum-rate by factors more than 1.24 and 1.36 for the case of $M=100$, respectively.
This is due to the fact that the ECF framework enables optimal transmit power of UEs which facilitates the exploitation of performance gain.
Furthermore, it can be seen from Fig. \ref{fig:1} that APS-PARA scheme is better than the PARA scheme.
This is contributed to the low IUI brought by the proposed AP selection.
Due to the effective noise variance which UEs' data need to tolerate decreases considerably, it is beneficial to utilize AP selection for improving the achievable sum-rate.
Besides, the computational complexity has also been reduced with AP selection.
For recovering UEs' original information, only $L$ integer linear combinations instead of $M$ need to be used in the power control.
Besides, when the number of APs is 60, compared with imperfect CSI estimated by MMSE estimation method \cite{bjornson2017massive} known at APs, the performance degradation caused by imperfect CSI is only 4\%.

Assuming that the power control is utilized at the CPU based on the large-scale fading, we need to replace ${g_{ml}}$ with ${\beta}_{ml}$ for solving the optimization problem (\ref{para_op}).
The PARA scheme using power control based on the large-scale fading is referred to as LSF-PARA scheme.
Fig. \ref{fig:2} shows the achievable sum-rate obtained with PARA, LSF-PARA, APS-PARA, and APS-LSF-PARA schemes against the number of APs.
As expected, the achievable sum-rate of all schemes improves as the number of APs increases, and applying AP selection does help enhance the performance.
Furthermore, the impact on achievable sum-rate of neglecting the small-scale fading is not critical, especially when the ratio of APs to UEs becomes large.
In particular, the performance gap due to ignoring the small-scale fading vanishes for $M=100$.
This is due to the property of channel hardening \cite{bjornson2017massive}.
As the number of antennas is sufficiently large, the variance of the channel gain reduces and the fading becomes almost as a deterministic channel.

\subsubsection{Successive Computation}
\begin{figure}[t!]
\centering
\includegraphics[width=3.5in]{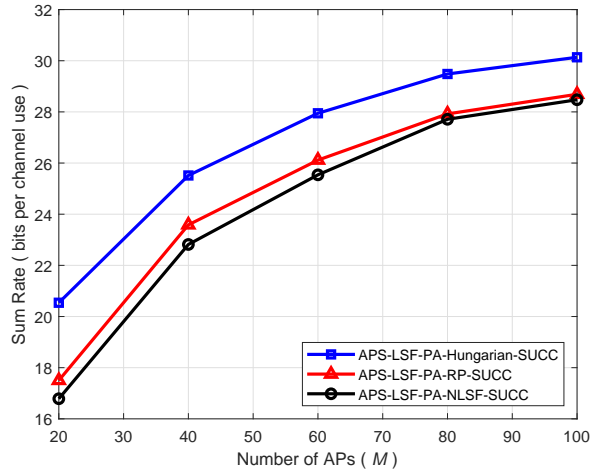}
\caption{Achievable sum-rate for APS-LSF-SUCC schemes with power allocation applying Hungarian algorithm, RP algorithm and NLSF algorithm for $L=10$ and ${P_{t}=200}$ mW.}
\label{fig:4}
\end{figure}

Next, we examine the performance of successive computation (SUCC) schemes. Let us denote the successive computation scheme based on large-scale fading applied with AP selection, power allocation, and Hungarian algorithm by APS-LSF-PA-Hungarian-SUCC.
Fig. \ref{fig:4} shows the performance of APS-LSF-PA-SUCC schemes with different algorithms on determining the decoding order of UEs, Hungarian algorithm, received-power-based (RP) algorithm, and channel-coefficient-based algorithm.
As we assume that the power control is employed at the CPU, which means that the channel coefficient is replaced with the large-scale fading coefficient, the APS-LSF-PA-SUCC scheme with searching the decoding order of UEs through 2-norm of large-scale fading coefficients is named as APS-LSF-PA-NLSF-SUCC scheme.
As shown in Fig. \ref{fig:4}, the APS-LSF-PA-Hungarian-SUCC scheme achieves the best result compared to other schemes. Furthermore, the performance of APS-LSF-PA-RP-SUCC scheme is similar to that of APS-LSF-PA-NLSF-SUCC.
This is due to the fact that the denominator of the second term in (\ref{para_sigema}) is several orders of magnitude smaller than the numerator, as the transmit power normalized by the noise power is huge.
Note that the effect of the first term in (\ref{para_sigema}) on effective noise variance is not significant.
Therefore, according to (\ref{para_sigema}), UEs with good channel state should be allocated with more transmit power for reducing the effective noise and finally the system can obtain a larger achievable sum-rate.
In this way, using RP and NLSF algorithms leads to the same result.

\begin{figure}[t!]
\centering
\includegraphics[width=3.5in]{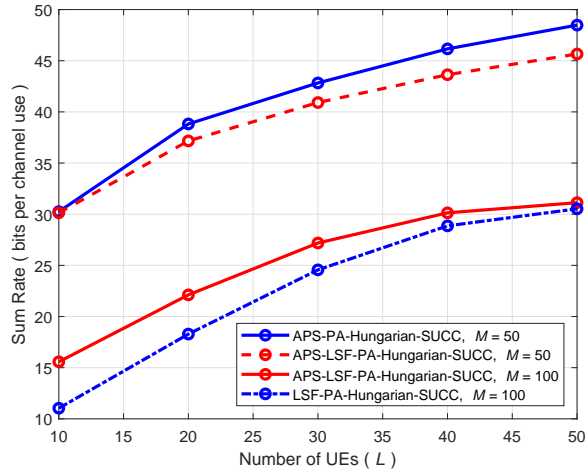}
\caption{Achievable sum-rate for APS-LSF-PA-Hungarian-SUCC and APS-PA-Hungarian-SUCC schemes for $M=50$ and $M=100$.}
\label{fig:5}
\end{figure}

In previous simulation results, we have shown that the instantaneous channel state information can help the parallel computation scheme to improve the achievable sum-rate while with higher complexity.
In successive computation, the conclusion is similar.
In Fig. \ref{fig:5}, we compare the achievable sum-rate of APS-PA-Hungarian-SUCC and APS-LSF-PA-Hungarian-SUCC schemes.
Although the performance gap induced by the replacement of channel coefficient becomes large along with the increase of the number of UEs, it is still very small.
At the same time, transmitting instantaneous channel state information yields a great growth load in fronthaul load.
Noted that the small-scale fading coefficient is only static during one coherence block while the large-scale fading coefficient stays constant for a duration of at least 40 small-scale fading coherence intervals \cite{ngo2017cell}.
Therefore, using the statistical channel state information works well for successive computation schemes.
Besides, the benefit of employing AP selection is also obvious in successive computation.
During searching the decoding order of combinations (\ref{RANK}), in the $m$th step we only need to calculate $L'-m$ times to find the minimal effective noise which increases the rank of the side information matrix.
Furthermore, we can use ${{\bf{u}}_1} \ldots ,{{\bf{u}}_{m - 1}}$ to eliminate certain symbols from the combination and thus remove the constraint on them.
For determining the total decoding order of combinations, the computational complexity is $\mathcal{O}\left( {{{\left( {L'} \right)}^2}} \right)$ while abandoning the selection needs $\mathcal{O}\left( {{{\left( {M'} \right)}^2}} \right)$.
Therefore, utilizing AP selection not only improves the performance but also decrease the computational complexity when the number of APs is larger than UEs.

\subsubsection{Comparison of centralized MMSE, ECF, CF, and MRC scheme}
\begin{figure}[t!]
\centering
\includegraphics[width=3.5in]{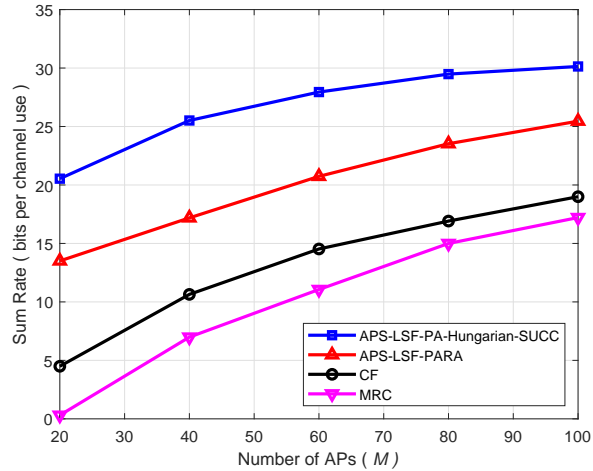}
\caption{Achievable sum-rate for APS-LSF-PA-Hungarian-SUCC scheme, APS-LSF-PARA scheme, CF scheme, and MRC scheme for $L=10$ and ${P_{t}=200}$ mW.}
\label{fig:8}
\end{figure}

\begin{figure}[t!]
\centering
\includegraphics[width=3.5in]{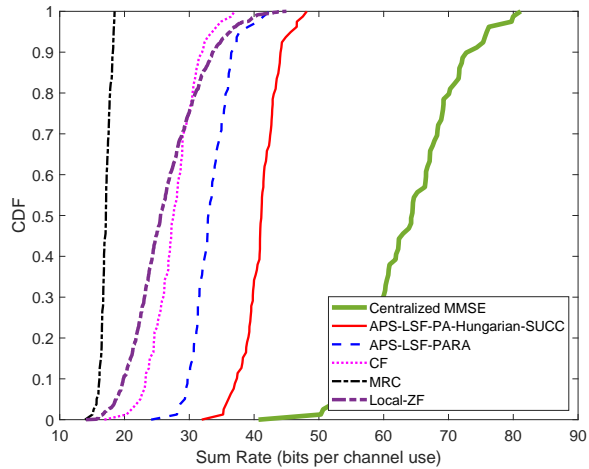}
\caption{CDFs of the achievable sum-rate for centralized MMSE, APS-LSF-PA-Hungarian-SUCC, APS-LSF-PARA, CF, Local ZF, and MRC schemes for $M=100$, $L=20$. }
\label{figCMMSE}
\end{figure}

In Fig. \ref{fig:8}, we compare the achievable sum-rate of APS-LSF-PA-Hungarian-SUCC, APS-LSF-PARA, CF, and MRC schemes.
MRC scheme is the simple linear strategy in cell-free massive MIMO, which has been widely used in previous works \cite{ngo2017cell}. In the uplink data transmission, the received signal at the $m$th AP can be expressed as
${{\bf{y}}_m} = \sum\nolimits_{l = 1}^L {{g_{ml}}} \sqrt {{{P_l}}} {{\bf{x}}_l} + {{\bf{z}}_m}$.
Then, the $m$th AP multiplies the received signal with the conjugate of its channel coefficient vector ${{{\bf{g}}}_m}$ and then forwards ${{\bf{y}}_m}{\bf{g}}_m^*$ to the CPU. The CPU combines signals from all $M$ APs. Therefore, the achievable rate of the $l$th UE is given by
\begin{equation}\label{MRC}
{R_{{\rm{mrc,}}l}} = {\log _2}\left( {1 + \frac{{P{{\left| {{\bf{g}}_l^H{{\bf{g}}_l}} \right|}^2}}}{{{{\left| {{\bf{g}}_l^H} \right|}^2} + {P_l}\sum\nolimits_{l \ne l'} {{{\left| {{\bf{g}}_{l'}^H{{\bf{g}}_l}} \right|}^2}} }}} \right).
\end{equation}
It is clear to see that the IUI limits the achievable sum-rate.
However, employing CF and ECF schemes can harness and even exploit the interference for cooperative gain, which leads to an increase in the achievable sum-rate.
This can be verified from Fig. \ref{fig:8}.
When the number of APs is not very large, which means the IUI affects the performance significantly, the advantage of applying CF and ECF schemes is self-evident.
For example, compared with MRC scheme, the achievable sum-rate of CF and ECF schemes improves by factors more than 1.5 and 2.5 when $M=100$, respectively.

Although utilizing the ECF framework can effectively improve the system performance, it is not the optimal choice for maximizing the achievable rate.
Fig. \ref{figCMMSE} shows the cumulative distribution function (CDF) of achievable sum-rate for centralized MMSE, APS-LSF-PA-Hungarian-SUCC, APS-LSF-PARA, CF, and MRC schemes with $M=100$, $L=20$.
From Fig. \ref{figCMMSE}, we can first observe that our parallel ECF scheme with power control method that solves (\ref{052802}) outperforms both CF and MRC.
Second, when comparing the ECF framework with local MR and zero-forcing (ZF) schemes using quantized signals under the same fronthaul limit, our proposed ECF schemes including parallel and successive computation have superior performance. Specifically, compared with the local ZF, applying the successive computation scheme leads to 60.4\% improvement in terms of the average achievable sum-rate.
Besides, the performance gap between the APS-LSF-PA-Hungarian-SUCC scheme and the centralized MMSE scheme is obvious. It is attributed to the fact that the centralized MMSE adopts the optimal combining scheme for maximizing the instantaneous signal-to-interference-and-noise ratio \cite{bjornson2017massive}.
Specifically, applying the centralized MMSE scheme leads to 55\% improvement in terms of average achievable sum-rate.
However, compared with centralized MMSE, ECF is also an efficient approach for fronthaul reduction and hence a largely achievable sum-rate still can be realized even if the fronthaul capacity is limited.
In particular, each AP decodes the received signal into the finite field by applying the equalization factor and then forwards an integer combination of the transmitted symbols of all UEs.
The cardinality of signals transmitted in the fronthaul link is the same as the cardinality of UEs original data, this is the theoretical minimum fronthaul load required to achieve lossless transmission \cite{huang2017compute}.
When the fronthaul capacity restricted as $R_0$, the actual achievable rate is $R = \min \left\{ {{R_0},{R_{\text{sum}}}} \right\}$ \cite{nazer2009structured}, where ${R_{\text{sum}}}$ represents the achievable sum-rate without considering the fronthaul load constraint.

\subsubsection{Trade-off between the performance and the complexity of ECF schemes}
\renewcommand\arraystretch{1.3}
\begin{table*}[tp]
  \centering
  \fontsize{9}{12}\selectfont
  \caption{The computational complexity and performance of various version of ECF framework}
  \label{tab:complexity}
    \begin{tabular}{ !{\vrule width1.2pt}  m{1.8cm}<{\centering} !{\vrule width1.2pt}  m{1.8cm}<{\centering} !{\vrule width1.2pt} m{3.0cm}<{\centering} !{\vrule width1.2pt}  m{4.8cm}<{\centering} !{\vrule width1.2pt}  m{1.8 cm}<{\centering} !{\vrule width1.2pt}  m{1.6cm}<{\centering} !{\vrule width1.pt}}

    \Xhline{1.2pt}
        \multicolumn{2}{|c|}{\multirow{2}{*}{ {\bf Schemes}  }}
        & \multicolumn{3}{|c|}{ {\bf Computation Complexity}  }
        & \multirow{2}{*}{ \tabincell{c}{\bf Sum rate \\ \bf [bits per \\ \bf channel use]}}\\
        \cline{3-5}
         \multicolumn{2}{|c|}{} & {AP Selection} & {\tabincell{c}{Power Optimization \\ (the number of feasibility problems)}} & {Searching Decoding Order}&
        \\\hline

    \Xhline{1.2pt}
        \multicolumn{2}{|c|}{ \tabincell{c}{Parallel \\Computation} }
        &
        ${\begin{array}{l}{\cal O}\left( {M' + M'{{\log }_2}\left( {M'} \right)} \right.\\\left. { + M'{{\left( {M' - 1} \right)}^3}} \right)\end{array}}$
        &
        $\begin{array}{l}\left\lfloor {\left( {{r_{\max }} \!+\! {r_{\max }} \!-\! {r_{sl}}\left\lfloor {\frac{{{r_{\max }} \!-\! {r_{\min }}}}{{{r_{sl}}}} \!\!-\!\! 1} \right\rfloor } \right)}\right.\\
        \left.\times \left\lfloor {\frac{{{r_{\max }} - {r_{\min }}}}{{{r_{sl}}}}} \right\rfloor /2{s_{sl}} \right\rfloor \left\lfloor {\frac{{{r_{\max }} - {r_{\min }}}}{{{r_{sl}}}}} \right\rfloor
        \end{array}$
        &
        {N/A}
        &
        19.57
        \\\hline

    \Xhline{1.2pt}
        \multirow{3}{*}{ \tabincell{c}{Successive\\Computation}  }

        &
        Using Received-power-based algorithm
        &
        $\begin{array}{l}{\cal O}\left( {M' + M'{{\log }_2}\left( {M'} \right)} \right.\\\left. { + M'{{\left( {M' - 1} \right)}^3}} \right)\end{array}$
        &
        $\begin{array}{l}\sum\limits_{m = 2}^{L'} {\frac{{\left( {L' - m + 2} \right)\left( {L' - m + 1} \right)}}{2}} \left[ {\frac{{{{\left( {L'} \right)}^3} - L'}}{3}} \right.\\ + \frac{{{{\left( {m - 1} \right)}^2}L' + \left( {m - 1} \right)L'}}{2} + {\left( {m - 1} \right)^3}L'\\\left. { + \left( {m - 1} \right){{\left( {L'} \right)}^2} + 2{{\left( {L'} \right)}^2}} \right]\end{array}$
        &
        ${\cal O}\left( {\frac{{\left( {L' + 1} \right)L'}}{2}} \right)$
        &
        23.58
        \\

        \cline{2-6}
        &
        Using Channel-coefficient-based algorithm
        &
        $\begin{array}{l}{\cal O}\left( {M' + M'{{\log }_2}\left( {M'} \right)} \right.\\\left. { + M'{{\left( {M' - 1} \right)}^3}} \right)\end{array}$
        &
        $\begin{array}{l}\sum\limits_{m = 2}^{L'} {\frac{{\left( {L' - m + 2} \right)\left( {L' - m + 1} \right)}}{2}} \left[ {\frac{{{{\left( {L'} \right)}^3} - L'}}{3}} \right.\\ + \frac{{{{\left( {m - 1} \right)}^2}L' + \left( {m - 1} \right)L'}}{2} + {\left( {m - 1} \right)^3}L'\\\left. { + \left( {m - 1} \right){{\left( {L'} \right)}^2} + 2{{\left( {L'} \right)}^2}} \right]\end{array}$
        &
        ${\cal O}\left( {\frac{{\left( {L' + 1} \right)L'}}{2}} \right)$
        &
        22.82
        \\

        \cline{2-6}
        &
        Using Hungarian algorithm
        &
        $\begin{array}{l}{\cal O}\left( {M' + M'{{\log }_2}\left( {M'} \right)} \right.\\\left. { + M'{{\left( {M' - 1} \right)}^3}} \right)\end{array}$
        &
        $\begin{array}{l}\sum\limits_{m = 2}^{L'} {\frac{{\left( {L' - m + 2} \right)\left( {L' - m + 1} \right)}}{2}} \left[ {\frac{{{{\left( {L'} \right)}^3} - L'}}{3}} \right.\\ + \frac{{{{\left( {m - 1} \right)}^2}L' + \left( {m - 1} \right)L'}}{2} + {\left( {m - 1} \right)^3}L'\\\left. { + \left( {m - 1} \right){{\left( {L'} \right)}^2} + 2{{\left( {L'} \right)}^2}} \right]\end{array}$
        &
        ${\cal O}\left( {{{\left( {L'} \right)}^3}} \right)$
        &
        25.51\\ \hline

    \Xhline{1.2pt}
    \end{tabular}
\end{table*}

In Table \ref{tab:complexity}, we summarize the performance in terms of sum-rate and the computational complexity of various versions of successive computation and parallel computation.
It can be observed that there is a trade-off between performance and computational complexity.
Specifically, the successive computation with the Hungarian algorithm for searching decoding order of UEs has the higher computational complexity and superior performance compared with the other two methods, i.e., received-power-based algorithm and channel-coefficient-based algorithm.

\subsubsection{Scalable Issue}
\begin{figure}[t!]
\centering
\includegraphics[width=3.5in]{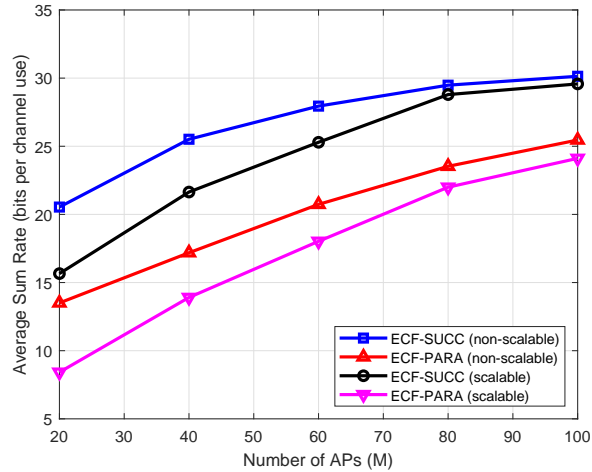}
\caption{Achievable sum-rate for the non scalable and scalable ECF-SUCC and ECF-PARA scheme for $L = 10$ and
$P_t = 200$ mW.}
\label{figscalable}
\end{figure}

In order to realize scalability \cite{bjornson2019scalable}, \cite{interdonato2019scalability}, our ECF framework needs to control the number of UEs each AP serves. Specifically, according to the large-scale-fading-based AP selection criterion proposed in \cite{ngo2017total}, APs are first selected to form UE-centric clusters for each UE.
Then, each AP sorts the UEs that need to be served according to the large-scale fading information, and then selects only the first several UEs with the best channel quality to serve.
Fig. \ref{figscalable} shows the performance comparison between the original non scalable ECF and the scalable ECF schemes. We can observe that the performance loss is small and decreases with the increase of the number of APs.

\section{Conclusions}\label{conclusion}
In this work, we investigate the achievable sum-rate of ECF framework for cell-free massive MIMO systems.
Two types of ECF framework including parallel computation and successive computation to improve the achievable sum-rate in cell-free massive MIMO are proposed.
An AP selection scheme is proposed to reduce the effective noise tolerance of UEs to further improve the performance and reduce the computation complexity.
We prove that the proposed power control algorithm for parallel computation and successive computation with AP selection can improve the achievable sum-rate significantly.
For obtaining better system performance, methods for determining the decoding order of combinations and UEs are also presented.
Numerical results show that compared with CF and MRC schemes, the ECF framework remarkably improves the achievable sum-rate of cell-free massive MIMO systems.





\end{document}